# An Improved Sphere-Packing Bound Targeting Codes of Short to Moderate Block Lengths and Applications

Gil Wiechman Igal Sason

Department of Electrical Engineering
Technion – Israel Institute of Technology
Haifa 32000, Israel
{igillw@tx, sason@ee}.technion.ac.il

***Abstract*— This paper derives an improved sphere-packing (ISP) bound targeting codes of short to moderate block lengths. We first review the 1967 sphere-packing (SP67) bound for discrete memoryless channels, and a recent improvement by Valembois and Fossorier. These concepts are used for the derivation of a new lower bound on the decoding error probability (referred to as the ISP bound) which is uniformly tighter than the SP67 bound and its recent improved version. Under a mild condition, the ISP bound is applicable to general memoryless channels, and some of its applications are exemplified. Its tightness is studied by comparing it with bounds on the ML decoding error probability. It is exemplified that the ISP bound suggests an interesting alternative to the 1959 sphere-packing (SP59) bound of Shannon for the Gaussian channel, especially for digital modulations of high spectral efficiency.**

## I. INTRODUCTION

The introduction of turbo-like codes which closely approach the Shannon capacity limit with moderate block lengths stirred up new interest in studying the limits of code performance as a function of the block length (see, e.g., [2], [5], [6], [8], [9], [12], [13], [14]).

The 1959 sphere-packing (SP59) bound of Shannon [10] serves for the evaluation of the performance limits of block codes whose transmission takes place over an AWGN channel. The bound is expressed in terms of the block length and rate of the code; it does not take into account the modulation used, but only assumes that the signals are of equal energy. This lower bound on the decoding error probability is often used as a reference for quantifying the sub-optimality of codes with their practical decoding algorithms; by comparing computer simulations for the performance obtained by turbo-like codes over a wide range of rates and block sizes, it was exemplified in the literature that the gap between the sphere-packing bounds and the performance of these codes under efficient iterative decoding algorithms can be reduced below 1 dB.

The 1967 sphere-packing (SP67) bound, derived by Shannon, Gallager and Berlekamp [11], provides a lower bound on the decoding error probability of block codes as a function of their block length and code rate, and it applies to arbitrary discrete memoryless channels. Like the random coding bound of Gallager [3], the SP67 bound decays to zero exponentially with the block length for all rates below the channel capacity. Further, the error exponent of the SP67 bound is tight at the portion of the rate region between the critical rate $(R_c)$ and the channel capacity; for this important rate region, the error exponents of the SP67 and the random coding bounds coincide [11, Part 1].

The SP67 bound appears to be loose for codes of small to moderate block lengths. This is due to the original focus in [11] on asymptotic analysis. In their paper [13], Valembois and Fossorier revisited the SP67 bound in order to improve its tightness for codes of short to moderate block lengths, and also to extend its validity to memoryless continuous-output channels (e.g., the binary-input AWGN channel). The motivation for their study in [13] was strengthened by the outstanding performance of codes defined on graphs even with moderate block lengths. The remarkable improvement of their bound over the classical SP67 bound was exemplified in [13], where it was shown that a new tightened version of the SP67 bound provides an interesting alternative to the SP59 bound [10].

In this work, we derive an improved sphere-packing bound (referred to as the ISP bound) which further enhances the tightness of the bounding technique in [11], especially for codes of short to moderate block lengths. Under a mild condition, the validity of this new bound is extended to general memoryless channels; it is applied here to M-ary PSK block coded modulation schemes whose transmission takes place over an AWGN channel. The tightness of the ISP bound is studied by comparing it with the random coding upper bound of Gallager [3], the tangential-sphere bound of Poltyrev ([4], [7]), and classical and recent sphere-packing bounds (see [10], [11], [13]). The tightness of the ISP bound for the Gaussian channel is also examined by calculating the regions of code lengths and rates for which this bound outperforms the SP59 bound and the capacity-limit bound (CLB). To this end, we present in [15, Section 4] (and [9]) a technique to perform the entire calculation of the SP59 bound in the logarithmic domain; this approach circumvents numerical difficulties, and facilitates an exact calculation of the SP59 bound for moderate to large block lengths without the need for the asymptotic approximations in [10].

The paper is structured as follows: Section II reviews the concepts used in the derivation of the SP67 bound [11, Part 1], and its recent improvements in [13] targeting codes of short to moderate block lengths. Section III introduces the ISP bound which further enhances the tightness of the bound in [13] and extends its validity for memoryless channels; the derivation of this bound relies on concepts and notation presented in Section II. Section IV provides numerical results which serve to compare the tightness of the ISP bound in Section III with the SP59 bound of Shannon [10] and the recent sphere-packing bound in [13]. The tightness of the ISP bound is exemplified in Section IV for M-ary phase-shift-keying (PSK) block coded modulation schemes whose transmission takes place over the AWGN channel. We conclude our discussion in Section V.

This paper presents in part the full paper version [15]. Another conference paper [9] which is derived from [15] considers a new algorithm for the efficient calculation of the SP59 bound.

## II. THE 1967 SPHERE-PACKING BOUND AND THE RECENT BOUND OF VALEMBOIS & FOSSORIER

Let us consider a block code $\mathcal{C}$ which consists of $M$ codewords each of length $N$, and denote its codewords by $\mathbf{x}_1, \ldots, \mathbf{x}_M$. Assume that $\mathcal{C}$ is transmitted over a discrete memoryless channel (DMC) and the decoding is performed by a list decoder; for each received sequence $\mathbf{y}$, the decoder outputs a list of at most $L$ integers belonging to the set $\{1, 2, \ldots, M\}$ which correspond to the indices of the codewords. A list decoding error is declared if the index of the transmitted codeword does not appear in the list. In [11], the authors derive a lower bound on the decoding error probability of an arbitrary block code with $M$ codewords of length $N$, and an arbitrary list decoding scheme whose size is limited to $L$. The particular case where $L = 1$ clearly provides a lower bound on the decoding error probability under ML decoding.

Let $\mathcal{Y}_m$ denote the set of output sequences $\mathbf{y}$ for which message $m$ is on the decoding list, and define $P_m(\mathbf{y}) \triangleq \Pr(\mathbf{y}|\mathbf{x}_m)$. The probability of list decoding error when message $m$ is sent is given by

$$P_{\text{e},m} = \sum_{y \in \mathcal{Y}_m^c} P_m(\mathbf{y}) \ . \tag{1}$$

For the block code and list decoder under consideration, let $P_{\text{e,max}}$ designate the maximal value of $P_{\text{e},m}$ where $m \in \{1, 2, \ldots, M\}$. Assuming that the codewords are equally likely to be transmitted, the average decoding error probability is given by

$$P_{\text{e}} = \frac{1}{M} \sum_{m=1}^{M} P_{\text{e},m}.$$

Referring to a list decoder of size at most $L$, the code rate (in nats per channel use) is defined as $R \triangleq \frac{\ln\left(\frac{M}{L}\right)}{N}$.

The derivation of the SP67 bound is divided into three main steps. The first step is the derivation of upper and lower bounds on the error probability of a code consisting of two codewords only. The authors prove in [11] the following theorem:

**Theorem II.1 (Upper and Lower Bounds on the Pairwise Error Probability).** [11, Theorem 5]: Let $P_1(\mathbf{y})$ and $P_2(\mathbf{y})$ be two probability assignments on a discrete set of sequences, $\mathcal{Y}_1$ and $\mathcal{Y}_2$ be disjoint decision regions for these sequences, $P_{\text{e},1}$ and $P_{\text{e},2}$ be given by (1), and let us assume that $P_1(\mathbf{y})P_2(\mathbf{y}) \neq 0$ for at least one sequence $\mathbf{y}$. Then, for all $s \in (0, 1)$

$$P_{\text{e},1} > \frac{1}{4} \exp\Bigl(\mu(s) - s\mu'(s) - s\sqrt{2\mu''(s)}\Bigr) \tag{2}$$

or

$$P_{\text{e},2} > \frac{1}{4} \exp\Bigl(\mu(s) + (1-s)\mu'(s) - (1-s)\sqrt{2\mu''(s)}\Bigr) \tag{3}$$

where

$$\mu(s) \triangleq \ln\Bigl(\sum_{\mathbf{y}} P_1(\mathbf{y})^{1-s} P_2(\mathbf{y})^s\Bigr) \qquad 0 < s < 1. \tag{4}$$

Furthermore, for an appropriate choice of $\mathcal{Y}_1$ and $\mathcal{Y}_2$

$$P_{\text{e},1} \leq \exp\Bigl(\mu(s) - s\mu'(s)\Bigr)$$

and

$$P_{\text{e},2} \leq \exp\Bigl(\mu(s) + (1-s)\mu'(s)\Bigr).$$

The function $\mu$ is non-positive and convex over the interval $(0, 1)$. The convexity of $\mu$ is strict unless $\frac{P_1(\mathbf{y})}{P_2(\mathbf{y})}$ is constant over all the sequences $\mathbf{y}$ for which $P_1(\mathbf{y})P_2(\mathbf{y}) \neq 0$. Moreover, the function $\mu$ is strictly negative over the interval $(0, 1)$ unless $P_1(\mathbf{y}) = P_2(\mathbf{y})$ for all $\mathbf{y}$.

The initial motivation given for Theorem II.1 is the calculation of bounds on the error probability of a two-word code. However, it is valid for any pair of probability assignments $P_1$ and $P_2$ and decision regions $\mathcal{Y}_1$ and $\mathcal{Y}_2$ which form a partitioning of the output vector space.

In the continuation of the derivation of the SP67 bound in [11], this theorem is used in order to keep control of the size of a decision region of a particular codeword without directly referring to other codewords. To this end, an arbitrary probability tilting measure $f_N$ is introduced in [11] over all $N$-length sequences of channel outputs, requiring that it is factorized in the form

$$f_N(\mathbf{y}) = \prod_{n=1}^{N} f(y_n) \tag{5}$$

for an arbitrary output sequence $\mathbf{y} = (y_1, \ldots, y_N)$; the size of the set $\mathcal{Y}_m$ is defined as

$$F(\mathcal{Y}_m) \triangleq \sum_{\mathbf{y} \in \mathcal{Y}_m} f_N(\mathbf{y}). \tag{6}$$

Next, [11] relies on Theorem II.1 in order to relate the conditional error probability $P_{\text{e},m}$ and $F(\mathcal{Y}_m)$ for fixed composition codes; this is done by associating $\Pr(\mathbf{y}|\mathbf{x}_m)$ and $f_N(\mathbf{y})$ with $P_1(\mathbf{y})$ and $P_2(\mathbf{y})$, respectively. Theorem II.1 is applied as described above to derive a parametric lower bound on the size of the decision region $\mathcal{Y}_m$ or the conditional error probability $P_{\text{e},m}$. Using a simple upper bound on the smallest size of the set $\mathcal{Y}_m$ where $m \in \{1, \ldots, M\}$, and upper bounding the conditional error probability of the corresponding codeword by $P_{\mathrm{e,max}}$, a lower bound on the maximal error probability is obtained. Next, the probability assignment $f \triangleq f_s$ is optimized in [11], so as to get the tightest (i.e., maximal) lower bound within this form while considering a code whose composition minimizes the bound (so that the bound holds for all fixed composition codes). A solution for this min-max problem, as provided in [11], leads to the following theorem which gives a lower bound on the maximal block error probability of an arbitrary fixed composition block code (for a more detailed review of these concepts, see [8, Section 5.3]).

**Theorem II.2 (Sphere-Packing Lower Bound on the Maximal Decoding Error Probability for Fixed Composition Codes).** [11, Theorem 6]: Let $\mathcal{C}$ be a *fixed composition code* of $M$ codewords and block length $N$. Assume that the transmission of $\mathcal{C}$ takes place over a DMC, and let $P(j|k)$ be the set of transition probabilities characterizing this channel (where $j \in \{1,\ldots,J\}$ and $k \in \{1,\ldots,K\}$ designate the channel input and output alphabets, respectively). For an arbitrary list decoder where the size of the list is limited to $L$, the *maximal error probability* ($P_{\mathrm{e,max}}$) satisfies

$$P_{\mathrm{e,max}} \geq \exp\left[-N\Big(E_{\mathrm{sp}}\Big(R - \frac{\ln 4}{N} - \varepsilon\Big) + \sqrt{\frac{8}{N}}\ln\Big(\frac{e}{\sqrt{P_{\min}}}\Big) + \frac{\ln 4}{N}\Big)\right]$$

where $R \triangleq \frac{\ln\left(\frac{M}{L}\right)}{N}$ is the rate of the code, $P_{\min}$ designates the smallest non-zero transition probability of the DMC, the parameter $\varepsilon$ is an arbitrarily small positive number, and the function $E_{\mathrm{sp}}$ is given by

$$E_{\mathrm{sp}}(R) \triangleq \sup_{\rho \geq 0}\big(E_0(\rho) - \rho R\big) \tag{7}$$

$$E_0(\rho) \triangleq \max_{\mathbf{q}} E_0(\rho,\mathbf{q}) \tag{8}$$

$$E_0(\rho,\mathbf{q}) \triangleq -\ln\left(\sum_{j=1}^{J}\Big[\sum_{k=1}^{K} q_k P(j|k)^{\frac{1}{1+\rho}}\Big]^{1+\rho}\right). \tag{9}$$

The maximum in the RHS of (8) is taken over all probability vectors $\mathbf{q} = (q_1,\ldots,q_K)$, i.e., over all $\mathbf{q}$ with $K$ non-negative components summing to 1.

The reason for considering fixed composition codes in [11] is that in general, the optimal probability distribution $f_s$ may depend on the composition of the codewords through the choice of the parameter $s$ in $(0,1)$ (see [11, p. 96]).

The next step in the derivation of the SP67 bound is the application of Theorem II.2 towards the derivation of a lower bound on the maximal error probability of an arbitrary block code. This is performed by lower bounding the maximal error probability of the code by the maximal error probability of its largest fixed composition subcode. Since the number of possible compositions is polynomial in the block length, one can lower bound the rate of the largest fixed composition subcode by $R - O\left(\frac{\ln N}{N}\right)$ where $R$ is the rate of the original code. Clearly, the rate loss caused by considering this subcode vanishes when the block length tends to infinity; however, it loosens of the bound for short to moderate length codes. Finally, the bound on the maximal error probability is transformed into a bound on the average error probability by considering an expurgated code which contains half of the codewords of the original code with the lowest decoding error probability. This finally leads to the SP67 bound [11].

**Theorem II.3 (The 1967 Sphere-Packing Bound for Discrete Memoryless Channels).** [11, Theorem 2]: Let $\mathcal{C}$ be an arbitrary block code whose transmission takes place over a DMC. Assume that the DMC is specified by the set of transition probabilities $P(j|k)$ where $k \in \{1,\ldots,K\}$ and $j \in \{1,\ldots,J\}$ designate the channel input and output alphabets, respectively. Assume that the code $\mathcal{C}$ forms a set of $M$ codewords of length $N$ (i.e., each codeword is a sequence of $N$ letters from the input alphabet), and consider an arbitrary list decoder where the size of the list is limited to $L$. Then, the *average decoding error probability* of the code $\mathcal{C}$ satisfies

$$P_{\mathrm{e}}(N,M,L) \geq \exp\left\{-N\Big[E_{\mathrm{sp}}\Big(R - O_1\Big(\frac{\ln N}{N}\Big)\Big) + O_2\Big(\frac{1}{\sqrt{N}}\Big)\Big]\right\}$$

where $R \triangleq \frac{\ln\left(\frac{M}{L}\right)}{N}$, the error exponent $E_{\mathrm{sp}}(R)$ is introduced in (7), and the terms

$$O_1\Big(\frac{\ln N}{N}\Big) = \frac{\ln 8}{N} + \frac{K \ln N}{N} \tag{10}$$
$$O_2\Big(\frac{1}{\sqrt{N}}\Big) = \sqrt{\frac{8}{N}}\,\ln\Big(\frac{e}{\sqrt{P_{\min}}}\Big) + \frac{\ln 8}{N}$$

scale like $\frac{\ln N}{N}$ and the inverse of the square root of $N$, respectively (hence, they vanish as we let $N$ tend to infinity), and $P_{\min}$ denotes the smallest non-zero transition probability of the DMC.

### *A. Improvements on the 1967 Sphere-Packing Bound Introduced in [13]*

In [13], Valembois and Fossorier revisit the derivation of the SP67 bound, focusing this time on codes of short to moderate block lengths. They present four modifications to the classical derivation in [11] which improve the pre-exponent of the SP67 bound. The new bound derived in [13] is also valid for memoryless channels with discrete input and continuous output (as opposed to the SP67 bound which is only valid for DMCs). It is applied to the binary-input AWGN channel, and it is also compared with the SP59 bound which is valid for any set of equal energy signals transmitted over the AWGN channel; this comparison shows that the recent bound in [13] provides an interesting alternative to the SP59 bound, especially for high code rates. In this section, we review the improvement suggested in [13] and present the resulting bound.

The first modification suggested in [13] is the addition of a free parameter in the derivation of the lower bound on the decoding error probability of two-word codes; this free parameter is used in conjunction with Chebychev's inequality, and it is later optimized in order to get the tightest bound within this form.

A second improvement presented in [13] is related to a simplification in [11] where the second derivative of the function $\mu$, as is defined in (4), is upper bounded by $\frac{e}{\sqrt{P_{\min}}}$. This bound results in no asymptotic loss, but it can loosen the bound for short to moderate code lengths. By using the exact value of $\mu''$ instead, the tightness of the resulting bound is further improved in [13]. This modification also makes the bound suitable to memoryless channels with continuous output, as it is no longer required that $P_{\min}$ is positive. It should be noted that this causes a small discrepancy in the derivation of the bound; the derivation of a lower bound on the error probability which is *uniform* over all fixed composition codes relies on finding the composition which minimizes the lower bound. This optimization problem is solved in [11] for the case where the upper bound on $\mu''$ is applied and the same composition is used [13], without checking whether it is still that minimizing composition. However, as we see in the next section, for a wide class of channels the value of the bound is independent of the code composition; therefore, the bound of Valembois and Fossorier [13, Theorem 7] (referred to as the VF bound) stays valid. This class of channels includes all memoryless binary-input output-symmetric (MBIOS) channels; in particular, it includes the binary symmetric channel (BSC), and the binary-input AWGN channel considered in [13].

A third improvement in [13] concerns the particular selection of the value of $\rho \geq 0$ which leads to the derivation of Theorem II.3. In [11], $\rho$ is set to be the value $\tilde{\rho}$ which minimizes the error exponent of the SP67 bound (i.e., the upper bound on the error exponent). This choice emphasizes the similarity between the error exponents of the SP67 lower bound and the Gallager random coding upper bound, hence proving that the error exponent of the SP67 bound is tight for all rates above the critical rate of the channel. In order to tighten the bound for the finite length case, [13] chooses the value of $\rho$ to be $\rho^*$ which provides the tightest possible lower bound on the decoding error probability. The asymptotic accuracy of the original SP67 bound implies that as the block length tends to infinity, $\tilde{\rho} \to \rho^*$; however, for codes of finite length, this observation tightens the bound with almost no penalty in the computational cost of the resulting bound.

The fourth observation made in [13] concerns the final stage in the derivation of the SP67 bound. In order to get a lower bound on the maximal error probability of an arbitrary block code, the derivation in [11] considers the maximal error probability a fixed composition subcode of the original code. In [11], a simple lower bound on the size of the largest fixed composition subcode is given; namely, the size of the largest fixed composition subcode is not less than the size of the entire code divided by the number of possible compositions, whose value cannot exceed $\binom{N+K-1}{K-1}$. To simplify the final expression of the SP67, [11] applies the upper bound $\binom{N+K-1}{K-1} \leq N^K$. Since this expression is polynomial is the block length $N$, there is no asymptotic loss to the error exponent. However, by using the exact expression for the number of possible compositions, the bound in [13] is tightened for codes of short to moderate block lengths. Applying these four modifications in [13] yields an improved lower bound on the decoding error probability of block codes transmitted over memoryless channels with finite input alphabets. As mentioned above, these modifications also extend the validity of the new bound to memoryless channels with discrete input and continuous output. However, the requirement of a finite input alphabet still remains, as it is required in order to apply the bound to arbitrary block codes, and not only to fixed composition codes. The VF bound [13] is given in the following theorem:

**Theorem II.4 (Improvement on the 1967 Sphere-Packing Bound for Discrete Memoryless Channels).** [13, Theorem 7]: Under the assumptions and notation used in Theorem II.3, the *average decoding error probability* satisfies

$$P_{\text{e}}(N, M, L) \geq \exp\Big\{-N\widetilde{E}_{\text{sp}}(R, N)\Big\}$$

where

$$\widetilde{E}_{\text{sp}}(R, N) \triangleq \sup_{x > \frac{\sqrt{2}}{2}} \Big\{E_0(\rho_x) - \rho_x \Big(R - O_1\Big(\frac{\ln N}{N}, x\Big)\Big) + O_2\Big(\frac{1}{\sqrt{N}}, x, \rho_x\Big)\Big\}$$

and

$$\begin{aligned}
R &\triangleq \frac{\ln\left(\frac{M}{L}\right)}{N} \\
O_1\Big(\frac{\ln N}{N}, x\Big) &\triangleq \frac{\ln 8}{N} + \frac{\ln \binom{N+K-1}{K-1}}{N} - \frac{\ln\left(2 - \frac{1}{x^2}\right)}{N} \qquad (11) \\
O_2\Big(\frac{1}{\sqrt{N}}, x, \rho\Big) &\triangleq x\sqrt{\frac{8}{N}\sum_{k=1}^{K} q_{k,\rho}\nu_k^{(2)}(\rho)} + \frac{\ln 8}{N} - \frac{\ln\left(2 - \frac{1}{x^2}\right)}{N} \\
\nu_k^{(1)}(\rho) &\triangleq \frac{\sum_{j=1}^{J} \beta_{j,k,\rho} \ln \frac{\beta_{j,k,\rho}}{P(j|k)}}{\sum_{j=1}^{J} \beta_{j,k,\rho}} \\
\nu_k^{(2)}(\rho) &\triangleq \frac{\sum_{j=1}^{J} \beta_{j,k,\rho} \ln^2 \frac{\beta_{j,k,\rho}}{P(j|k)}}{\sum_{j=1}^{J} \beta_{j,k,\rho}} - \left[\nu_k^{(1)}(\rho)\right]^2 \\
\beta_{j,k,\rho} &\triangleq P(j|k)^{\frac{1}{1+\rho}} \cdot \left(\sum_{k'} q_{k',\rho} P(j|k')^{\frac{1}{1+\rho}}\right)^{\rho}
\end{aligned}$$

where $\mathbf{q}_\rho \triangleq (q_{1,\rho}, \ldots, q_{K,\rho})$ designates the input distribution which maximizes $E_0(\rho, \mathbf{q})$ in (8), and the parameter $\rho_x$ is determined by solving the equation

$$R - O_1\Big(\frac{\ln N}{N}, x\Big) = -\frac{1}{\rho_x}\sum_k q_{k,\rho_x}\nu_k^{(1)}(\rho_x) + \frac{x}{\rho_x}\sqrt{\frac{2}{N}\sum_{k=1}^{K} q_{k,\rho}\nu_k^{(2)}(\rho)}.$$

## III. AN IMPROVED SPHERE-PACKING BOUND

In this section, we derive an improved lower bound on the decoding error probability which utilizes the sphere-packing bounding technique. This bound is referred to as the improved sphere-packing (ISP) bound, and its validity is extended to a wide class of discrete-time memoryless channels.

To keep the notation simple, we derive the ISP bound under the assumption that the communication takes place over a DMC. This assumption allows us to follow the first steps of the proof of the SP67 bound in [11]. However, the derivation of the bound is justified later for a wider class of memoryless channels with discrete or continuous input and output alphabets. Some remarks are given at the end of the derivation.

We start our analysis by following the derivation of the SP67 bound, as given in [11], where we take advantage of the improvements suggested in [13]. We show that under a mild condition on memoryless communication channels, the derivation of the sphere-packing bound can be modified so that the intermediate step of bounding the maximal error probability for fixed composition codes can be skipped. This allows the tightening of the sphere-packing bound, and also the extension of its validity to the case where the channel input as well as the channel output are continuous. We begin the derivation by using the modified lower bound on the decoding error probability for a code book of two codewords, as presented in [13]. The novelty in the continuation is by moving directly to the derivation of the sphere-packing bound for a general block code, assuming a list decoder of size $L$, without the need to derive the bound first for fixed composition codes (thus differing from the derivation of the sphere-packing bounds in [11], [13]).

*a) Decoding Error Probability for Two Codewords:* We start the analysis by considering a lower bound on the decoding error probability of a codebook of two codewords, $\mathbf{x}_1$ and $\mathbf{x}_2$, whose transmission takes place over a DMC. This bound simply follows the idea in [13] and the analysis in [11]. The reader is referred to derivation of this bound in [15, Section 3.1.1].

Let $P_{\mathrm{e},1}$ and $P_{\mathrm{e},2}$ designate the decoding error probabilities of this code, conditioned on the transmission of the first and second codewords, respectively. The analysis shows that for every $s \in (0,1)$ and $x > \frac{\sqrt{2}}{2}$

$$P_{\mathrm{e},1} > \left(\frac{1}{2} - \frac{1}{4x^2}\right) \exp\Big(\mu(s) - s\mu'(s) - s\,x\,\sqrt{2\mu''(s)}\Big) \quad (12)$$

or

$$P_{\mathrm{e},2} > \left(\frac{1}{2} - \frac{1}{4x^2}\right) \exp\Big(\mu(s) + (1-s)\mu'(s) - (1-s)\,x\,\sqrt{2\mu''(s)}\Big). \quad (13)$$

*b) Lower Bound on the Decoding Error Probability of General Block Codes:* Let us now consider a block code $\mathcal{C}$ of length $N$ with $M$ codewords, denoted by $\{\mathbf{x}_m\}_{m=1}^M$; assume that the transmission takes place over a DMC with transition probabilities $P(j|k)$, where $k \in \{1,\ldots,K\}$ and $j \in \{1,\ldots,J\}$ designate the channel input and output alphabets, respectively. To this end, we rely on the result of the previous section which is valid for any pair of probability measures ($P_1$ and $P_2$). Let $f_N$ be an arbitrary probability measure defined over the set of length-$N$ sequences of the channel output, and which can be factorized as in (5). We refer to the pair of probability measures given by

$$P_1(\mathbf{y}) \triangleq \Pr(\mathbf{y}|\mathbf{x}_m), \quad P_2(\mathbf{y}) \triangleq f_N(\mathbf{y}) \quad (14)$$

where $\mathbf{x}_m$ is an arbitrary codeword of the code $\mathcal{C}$. Let $\mathcal{Y}_m$ be the decision region of the codeword $\mathbf{x}_m$, and let its size be defined as in (6). By combining (4) and (14), we define

$$\mu(s,m,f_N) \triangleq \ln\left(\sum_{\mathbf{y}} \Pr(\mathbf{y}|\mathbf{x}_m)^{1-s} f_N(\mathbf{y})^s\right), \quad 0 < s < 1. \quad (15)$$

By associating $\mathcal{Y}_m$ and $\mathcal{Y}_m^c$ with the two decision regions for the probability measures $P_1$ and $P_2$, respectively, we obtain from (12), (13) and the above setting that

$$P_{\mathrm{e},m} > \left(\frac{1}{2} - \frac{1}{4x^2}\right) \exp\Big(\mu(s,m,f_N) - s\mu'(s,m,f_N) - s\,x\,\sqrt{2\mu''(s,m,f_N)}\Big) \quad (16)$$

or

$$F(\mathcal{Y}_m) > \left(\frac{1}{2} - \frac{1}{4x^2}\right) \exp\Big(\mu(s,m,f_N) + (1-s)\mu'(s,m,f_N) - (1-s)\,x\,\sqrt{2\mu''(s,m,f_N)}\Big) \quad (17)$$

where $x > \frac{\sqrt{2}}{2}$.

Let us denote by $q_k^m$ the fraction of appearances of the letter $k$ in the codeword $\mathbf{x}_m$. By assumption, the communication channel is memoryless and the function $f_N$ is a probability measure which is factorized according to (5). Hence, for every $m \in \{1,2,\ldots,M\}$, the function $\mu(s,m,f_N)$ in (15) is expressible in the form

$$\mu(s,m,f_N) = N \sum_{k=1}^{K} q_k^m \mu_k(s,f) \quad (18)$$

where

$$\mu_k(s,f) \triangleq \ln\left(\sum_{j=1}^{J} P(j|k)^{1-s} f(j)^s\right), \quad 0 < s < 1. \quad (19)$$

Substituting (18) in (16) and (17), then for every $s \in (0,1)$

$$P_{\mathrm{e},m} > \left(\frac{1}{2} - \frac{1}{4x^2}\right) \exp\left\{N\left(\sum_k q_k^m \big(\mu_k(s,f) - s\mu_k'(s,f)\big) - s\,x\,\sqrt{\frac{2\sum_k q_k^m \mu_k''(s,f)}{N}}\right)\right\} \quad (20)$$

or

$$F(\mathcal{Y}_m) > \left(\frac{1}{2} - \frac{1}{4x^2}\right) \exp\left\{N\left(\sum_k q_k^m \big(\mu_k(s,f) + (1-s)\mu_k'(s,f)\big) - (1-s)\,x\,\sqrt{\frac{2\sum_k q_k^m \mu_k''(s,f)}{N}}\right)\right\}. \quad (21)$$

For $s \in (0,1)$, we choose the function $f$ to be $f_s$, as given in [11, Eqs. (4.18)-(4.20)]. Namely, for $0<s<1$, let $\mathbf{q}_s = \{q_{1,s}, \ldots, q_{K,s}\}$ satisfy the inequalities

$$\sum_j P(j|k)^{1-s} \alpha_{j,s}^{\frac{s}{1-s}} \geq \sum_j \alpha_{j,s}^{\frac{1}{1-s}} \; ; \quad \forall k \tag{22}$$

where

$$\alpha_{j,s} \triangleq \sum_{k=1}^{K} q_{k,s} P(j|k)^{1-s}. \tag{23}$$

The function $f_s$ is given by

$$f_s(j) = \frac{\alpha_{j,s}^{\frac{1}{1-s}}}{\sum_{j'=1}^{J} \alpha_{j',s}^{\frac{1}{1-s}}}, \quad j \in \{1, \ldots, J\}. \tag{24}$$

Note that the input distribution $\mathbf{q}_s$ is *independent of the code* $\mathcal{C}$, as it only depends on the channel statistics. By multiplying both sides of (22) by $q_{k,s}$ and summing over $k$ (where $\sum_k q_{k,s} = 1$), we get

$$\sum_k \left\{ q_{k,s} \sum_j P(j|k)^{1-s} \alpha_{j,s}^{\frac{s}{1-s}} \right\} \geq \sum_j \alpha_{j,s}^{\frac{1}{1-s}}.$$

Examining the LHS of the above inequality gives

$$\begin{aligned} &\sum_k \left\{ q_{k,s} \sum_j P(j|k)^{1-s} \alpha_{j,s}^{\frac{s}{1-s}} \right\} \\ &= \sum_j \left\{ \alpha_{j,s}^{\frac{s}{1-s}} \sum_k q_{k,s} P(j|k)^{1-s} \right\} \\ &= \sum_j \alpha_{j,s}^{\frac{1}{1-s}} \end{aligned} \tag{25}$$

where the last equality follows from (23). Let us now assume that for every $0<s<1$, the support of $\mathbf{q}_s$ consists of the entire input alphabet. By our assumption, $q_{k,s} \neq 0$ for any $k$, thus by combining (22) and (25), we obtain that for all values of $k$

$$\sum_j P(j|k)^{1-s} \alpha_{j,s}^{\frac{s}{1-s}} = \sum_j \alpha_{j,s}^{\frac{1}{1-s}}. \tag{26}$$

Note that this equality holds for all values of $k$ due to our assumption that $q_{k,s} \neq 0$ for all $k$; otherwise, this equality may not hold for those values of $k$ for which $q_{k,s}$ is zero. From (19), since $f$ in general is allowed to depend on the parameter $s$ (as we examine the validity of the bound for any individual value of $s \in (0,1)$), we get

$$\begin{aligned} &\mu_k(s, f_s) \\ &= \ln\Big(\sum_j P(j|k)^{1-s} f_s(j)^s\Big) \\ &\stackrel{(a)}{=} \ln\Big(\sum_j P(j|k)^{1-s} \alpha_{j,s}^{\frac{s}{1-s}}\Big) - s \ln\Big(\sum_j \alpha_{j,s}^{\frac{1}{1-s}}\Big) \\ &\stackrel{(b)}{=} (1-s) \ln\Big(\sum_j \alpha_{j,s}^{\frac{1}{1-s}}\Big) \\ &\stackrel{(c)}{=} (1-s) \ln\left(\sum_j \left[\sum_k q_{k,s} P(j|k)^{1-s}\right]^{\frac{1}{1-s}}\right) \end{aligned} \tag{27}$$

where $(a)$ follows from the definition of $f_s$ in (23) and (24), $(b)$ follows from (26), and $(c)$ follows from (23). Under the setting $s = \frac{\rho}{1+\rho}$, since the conditions on $\mathbf{q}_s$ in (22) are identical to the conditions on the input distribution $\mathbf{q} = \mathbf{q}_s$ which maximizes $E_0(\frac{s}{1-s}, \mathbf{q})$ as stated in [3, Theorem 4], then

$$\begin{aligned} &\mu_k(s, f_s) \\ &= (1-s) \ln\left(\sum_j \left[\sum_k q_{k,s} P(j|k)^{\frac{1}{1+\frac{s}{1-s}}}\right]^{1+\frac{s}{1-s}}\right) \\ &= -(1-s)\, E_0\left(\frac{s}{1-s}, \mathbf{q}_s\right) \\ &= -(1-s)\, E_0\left(\frac{s}{1-s}\right), \quad 0<s<1 \end{aligned} \tag{28}$$

where $E_0$ is given by (8). From (28), since $\mu_k$ is independent of $k$ (let its common value for all $k$ be denoted by $\mu_0(s, f_s)$), then from (20) and (21), it follows that for $0<s<1$

$$P_{\mathrm{e},m} > \left(\frac{1}{2} - \frac{1}{4x^2}\right) \exp\left\{ N \left(\mu_0(s,f_s) - s\mu_0'(s,f_s) - s\,x \sqrt{\frac{2\mu_0''(s,f_s)}{N}}\right)\right\} \tag{29}$$

or

$$F_s(\mathcal{Y}_m) > \left(\frac{1}{2} - \frac{1}{4x^2}\right) \exp\left\{ N \left(\mu_0(s,f_s) + (1-s)\mu_0'(s,f_s) - (1-s)\,x \sqrt{\frac{2\mu_0''(s,f_s)}{N}}\right)\right\} \tag{30}$$

where $F_s(\mathcal{Y}_m) \triangleq \sum_{\mathbf{y} \in \mathcal{Y}_m} f_{s,N}(\mathbf{y})$. Similarly to [11], we relate $F_s(\mathcal{Y}_m)$ to the number of codewords $M$ and the size of the decoding list $L$ by observing that

$$\sum_{m=1}^{M} F_s(\mathcal{Y}_m) = \sum_{m=1}^{M} \sum_{\mathbf{y} \in \mathcal{Y}_m} f_{s,N}(\mathbf{y}) \leq L.$$

The last inequality holds since each $\mathbf{y} \in \mathcal{Y}^N$ appears in at most $L$ subsets $\{\mathcal{Y}_m\}_{m=1}^M$ and also $\sum_{\mathbf{y}} f_{s,N}(\mathbf{y}) = 1$. It follows that for each $s \in (0,1)$, there exists an index $m_s \in \{1, 2, \ldots, M\}$ such that $F_s(\mathcal{Y}_{m_s}) \leq \frac{L}{M}$. Substituting this in (29) and (30), and upper bounding $P_{\mathrm{e},m_s}$ by the maximum over $m$ of $P_{\mathrm{e},m}$ (this maximal error probability is denoted by $P_{\mathrm{e,max}}$) implies that for $0<s<1$

$$P_{\mathrm{e,max}} > \left(\frac{1}{2} - \frac{1}{4x^2}\right) \exp\left\{ N \left(\mu_0(s,f_s) - s\mu_0'(s,f_s) - s\,x \sqrt{\frac{2\mu_0''(s,f_s)}{N}}\right)\right\} \tag{31}$$

or

$$\frac{L}{M} > \left(\frac{1}{2} - \frac{1}{4x^2}\right) \exp\left\{ N \left(\mu_0(s,f_s) + (1-s)\mu_0'(s,f_s) - (1-s)\,x \sqrt{\frac{2\mu_0''(s,f_s)}{N}}\right)\right\}. \tag{32}$$

A lower bound on the maximum error probability can be obtained from (31) by substituting any value of $s \in (0,1)$ for which the inequality in (32) does not hold. In particular we choose a value $s = s_x$ such that the inequality in (32) is replaced by equality, i.e.,

$$\begin{aligned}\frac{L}{M} &= \exp(-NR) \\ &= \left(\frac{1}{2} - \frac{1}{4x^2}\right) \exp\Bigg\{N\Big(\mu_0(s_x, f_{s_x}) + (1-s_x)\,\mu_0'(s_x, f_{s_x}) \\ &\qquad -(1-s_x)\,x\,\sqrt{\frac{2\mu_0''(s_x, f_{s_x})}{N}}\Big)\Bigg\} \quad (33)\end{aligned}$$

where $R \triangleq \frac{\ln\left(\frac{M}{L}\right)}{N}$ designates the code rate in nats per channel use. Note that the existence of a solution $s = s_x$ to equation (33) can be demonstrated in a similar way to the arguments in [11, Eqs. (4.28)–(4.35)] for the non-trivial case where the sphere-packing bound does not reduce to the trivial inequality $P_{\mathrm{e,max}} \geq 0$. This particular value of $s$ is chosen since for large enough $N$, the RHS of (31) is monotonically decreasing while the RHS of (32) is monotonically increasing for $s \in (0,1)$; thus, this choice is optimal for large enough $N$. The choice of $s = s_x$ also allows to get a simpler representation of the bound on $P_{\mathrm{e,max}}$. Rearranging equation (33) gives

$$\begin{aligned}\mu_0'(s_x, f_{s_x}) &= -\frac{1}{1-s_x}\left[R + \mu_0(s_x, f_{s_x}) + \frac{1}{N}\,\ln\left(\frac{1}{2} - \frac{1}{4x^2}\right)\right] \\ &\qquad + x\sqrt{\frac{2\mu_0''(s_x, f_{s_x})}{N}}.\end{aligned}$$

Substituting $s = s_x$ and the last equality into (31) gives

$$\begin{aligned}P_{\mathrm{e,max}} > \exp\Bigg\{N\Bigg(&\frac{\mu_0(s_x, f_{s_x})}{1-s_x} + \frac{s_x}{1-s_x}\left(R + \frac{\ln\left(\frac{1}{2} - \frac{1}{4x^2}\right)}{N}\right) \\ &- s_x\,x\sqrt{\frac{8\mu_0''(s_x, f_{s_x})}{N}} + \frac{\ln\left(\frac{1}{2} - \frac{1}{4x^2}\right)}{N}\Bigg)\Bigg\}.\end{aligned}$$

By applying (28) and defining $\rho_x \triangleq \frac{s_x}{1-s_x}$, we get

$$\begin{aligned}P_{\mathrm{e,max}} > \exp\Bigg\{-N\Bigg(&E_0(\rho_x) - \rho_x\left[R - \frac{\ln 4}{N} + \frac{\ln\left(2 - \frac{1}{x^2}\right)}{N}\right] \\ &+ s_x\,x\,\sqrt{\frac{8\mu_0''(s_x, f_{s_x})}{N}} + \frac{\ln 4}{N} - \frac{\ln\left(2 - \frac{1}{x^2}\right)}{N}\Bigg)\Bigg\}.\end{aligned}$$

Note that the above lower bound on the maximal error probability holds for an arbitrary block code of length $N$ and rate $R$. The selection of $\rho_x$ is similar to the selection in [13].

In order to transform the lower bound on the maximum error probability into a lower bound on the average error probability, we expurgate the original block code. In this standard approach, we look at the expurgated code which is comprised of the $\frac{M}{2}$ codewords with the lowest error probability. The average error probability of the original code is not below one-half of the maximal word error probability of the expurgated code. Since the rate of the expurgated code is $R' = R - \frac{\ln 2}{N}$ nats per channel use (the reduction in the rate by $\frac{\ln 2}{N}$ follows from reducing the size of the code by one-half), we get a lower bound on the average error probability of the original code which reads

$$\begin{aligned}P_{\mathrm{e}} > \exp\Bigg\{-N\Bigg(&E_0(\rho_x) - \rho_x\left[R - \frac{\ln 8}{N} + \frac{\ln\left(2 - \frac{1}{x^2}\right)}{N}\right] \\ &+ s_x\,x\,\sqrt{\frac{8\mu_0''(s_x, f_{s_x})}{N}} + \frac{\ln 8}{N} - \frac{\ln\left(2 - \frac{1}{x^2}\right)}{N}\Bigg)\Bigg\}\end{aligned}$$

where $R' - \frac{\ln 4}{N}$ is replaced in the RHS of the bound above by $R - \frac{\ln 8}{N}$, $\rho_x = \frac{s_x}{1-s_x}$ and $s_x \in (0,1)$ is implicitly given as a solution of the equation

$$\begin{aligned}R - \frac{\ln 8}{N} + \frac{\ln\left(2 - \frac{1}{x^2}\right)}{N} &= -\mu_0(s_x, f_{s_x}) - (1-s_x)\mu_0'(s_x, f_{s_x}) \\ &\qquad + (1-s_x)\,x\,\sqrt{\frac{2\mu_0''(s_x, f_{s_x})}{N}}.\end{aligned}$$

Finally, we optimize over the parameter $x \in (\frac{\sqrt{2}}{2}, \infty)$ in order to get the tightest lowest bound of this form.

The derivation above only relies on the fact that the channel is memoryless, but does not rely on the fact that the input or output alphabets are discrete. As mentioned in Section II-A, the original derivation of the SP67 bound in [11] relies on the fact that the input and output alphabets are finite in order to bound the second derivative of $\mu$ by $\frac{e}{\sqrt{P_{\min}}}$, where $P_{\min}$ designates the smallest non-zero transition probability of the channel. This requirement was relaxed in [13] to the requirement that only the input alphabet is finite; to this end, the second derivative of the function $\mu$ is calculated, thus the above upper bound on this second derivative is replaced by its exact value. However, the requirement for a finite input alphabet remains in [13] due to the fact that the derivation still relies on considering a fixed composition subcode of the original code, and therefore requires that the number of possible compositions for a given length $N$ is finite. The derivation in this section circumvents the use of fixed composition codes, and as a by-product, it also relaxes the requirement of a finite input alphabet. The validity of the derivation for memoryless continuous-input and continuous-output channels is provided in the continuation (see Remark III.4). This leads to the following theorem, which provides an improved sphere-packing lower bound on the error probability of block codes transmitted over memoryless channels.

**Theorem III.1 (An Improved Sphere-Packing (ISP) Bound for Memoryless Channels).** Let $\mathcal{C}$ be an arbitrary block code consisting of $M$ codewords, each of length $N$. Assume that $\mathcal{C}$ is transmitted over a memoryless channel which is specified by the transition probabilities (or densities) $P(j|k)$ where $k \in \mathcal{K}$ and $j \in \mathcal{J}$ designate the channel input and output alphabets, respectively. Assume an arbitrary list decoder where the size of the list is limited to $L$. If the support of $\mathbf{q}_s$ which satisfies the inequalities in (22) consists of the entire input alphabet for all $0 < s < 1$, then the *average decoding error probability* satisfies

$$P_{\text{e}}(N, M, L) \geq \exp\Big\{-N\widetilde{E}_{\text{sp}}(R, N)\Big\}$$

where

$$\widetilde{E}_{\text{sp}}(R, N) \triangleq \sup_{x > \frac{\sqrt{2}}{2}} \left\{E_0(\rho_x) - \rho_x \left(R - O_1\Big(\frac{1}{N}, x\Big)\right) + O_2\Big(\frac{1}{\sqrt{N}}, x, \rho_x\Big)\right\} \tag{34}$$

the function $E_0$ is introduced in (8), $R = \frac{1}{N}\ \ln\Big(\frac{M}{L}\Big)$, and

$$O_1\Big(\frac{1}{N}, x\Big) \triangleq \frac{\ln 8}{N} - \frac{\ln\Big(2 - \frac{1}{x^2}\Big)}{N} \tag{35}$$

$$O_2\Big(\frac{1}{\sqrt{N}}, x, \rho\Big) \triangleq s(\rho)\ x\ \sqrt{\frac{8}{N}\mu_0''\big(s(\rho), f_{s(\rho)}\big)} + \frac{\ln 8}{N} - \frac{\ln\Big(2 - \frac{1}{x^2}\Big)}{N}. \tag{36}$$

Here, $s(\rho) \triangleq \frac{\rho}{1+\rho}$, the non-negative parameter $\rho = \rho_x$ in the RHS of (34) is determined by solving the equation

$$R - O_1\Big(\frac{1}{N}, x\Big) = -\mu_0\big(s(\rho), f_{s(\rho)}\big) - \big(1 - s(\rho)\big)\mu_0'\big(s(\rho), f_{s(\rho)}\big) + \big(1 - s(\rho)\big)\ x\ \sqrt{\frac{2\mu_0''\big(s(\rho), f_{s(\rho)}\big)}{N}} \tag{37}$$

and the functions $\mu_0(s, f)$ and $f_s$ are defined in (19) and (24), respectively.

**Remark III.1.** The requirement that the support of the input distribution $\mathbf{q}_s$ which maximizes the sphere-packing error exponent consists of the entire input alphabet for all $s \in (0, 1)$ is crucial for the derivation of the ISP bound. It is basically what in essence makes the difference between the VF bound in [13] and the ISP bound here. Though it was commented by Shannon, Gallager and Berlekamp [11] that $\mu_k(s, f_s)$ in (27) is independent of $k$ for all letters $k$ for which $q_{k,s} > 0$, the implication of this observation in order to tighten the bound for finite length codes was not considered (note that [11] is focused on asymptotic analysis of error exponents). The novelty of the ISP bound is the use of this property for tightening the VF bound by skipping the intermediate step which is related to fixed composition codes since the requirement above allows us to represent the bound in (29) and (30) independently of the composition $\mathbf{q}^m$ of the codeword $\mathbf{x}_m$. In case, however, that the support of the input distribution $\mathbf{q}_s$ does not satisfy the above condition, the intermediate step of the derivation of the bound for fixed composition codes as in [11], [13] is required. This mutual dependency does not allow us in general to complete the proof for the general case.

**Remark III.2.** In light of the previous remark, the ISP bound differs from the VF bound [13] (see Theorem II.4) in the sense that the term $\frac{\log\binom{N+K-1}{K-1}}{N}$ is removed from $O_1(\frac{\ln N}{N}, x)$. Therefore, the shift in the rate of the error exponent of the ISP bound behaves asymptotically like $O_1\left(\frac{1}{N}\right)$ instead of $O_1\left(\frac{\ln N}{N}\right)$ (see (10), (11) and (35)). This difference indicates a tightening of the pre-exponent of the ISP bound (as compared to the SP67 and VF bounds) which is expected to be especially pronounced for small to moderate block lengths and when the size of the channel input alphabet is increased.

**Remark III.3.** The rate loss as a result of the expurgation of the code by removing half of the codewords with the largest error probability was ignored in [13]. The term $\frac{\ln 4}{N}$, as it appears in the term $O_1(\frac{\ln N}{N}, x)$ of [13, Theorem 7], should be therefore replaced by $\frac{\ln 8}{N}$ (see (35)).

**Remark III.4.** Under the mild condition discussed in Remark III.1, the ISP bound is also applicable to continuous-input memoryless channels. In contrast to (11) which depends on the size of the channel input alphabet ($K$)) and requires it to be finite, the parallel expression in (35) which corresponds to the ISP bound is not subject to this requirement. This inherent difference stems from Remark III.1. When the ISP bound is applied to a continuous-input memoryless channel, the distribution of the channel input, as used for the derivation of the bound for a DMC, is replaced by the probability density function of the continuous channel input. Similarly, the transition probability of a DMC is replaced by a transition density function for a memoryless channel with continuous input or output alphabets, and the sums are replaced by integrals. Note that these densities may include Dirac delta functions which appear at the points where the corresponding input distribution or the transition density function of the channel are discontinuous.

**Remark III.5.** Let $\alpha$ designate the fraction of codewords with the smallest error probability in the expurgated code (where $0 < \alpha < 1$). As discussed in [15, Discussion 3.1], for high block error probabilities, a rather small value of $\alpha$ improves the tightness of the bound. On the other hand, for low block error probabilities, larger values of $\alpha$ give more appealing results (note that, however, the bound is useless for $\alpha \to 1^-$). The reader is referred to [15, Discussion 3.1] for a discussion on this phenomenon, though it was exemplified that the choice $\alpha = \frac{1}{2}$ provides a rather tight bound uniformly.

## IV. NUMERICAL RESULTS

The ISP bound in Section III is particularized here to M-ary PSK block coded modulation schemes whose transmission takes place over the AWGN channel, and where the received signals are coherently detected. The closed form expressions for the function $\mu_0$ and its derivatives (see [15, Appendix A]) are useful for the calculation of the VF bound [13] as well. The SP59 bound [10] provides a lower bound on the decoding error probability for the considered case, since the modulated signals have equal energy and are transmitted over the AWGN channel. In the following, we exemplify the use of these lower bounds. They are also compared to the random-coding upper bound of Gallager [3], and the tangential-sphere upper bound (TSB) [7] when applied to random block codes. This serves for the study of the tightness of the ISP bound, as compared to other upper and lower bounds. The numerical results shown in this section indicate that the recent variants of the SP67 bound provide an interesting alternative to the SP59 bound which is commonly used in the literature as an ultimate measure of performance for codes transmitted over the AWGN channel (see, e.g., [2], [5], [6], [8], [12], [13], [14]). The advantage of the ISP bound over the VF bound in [13] is also exemplified in this section.

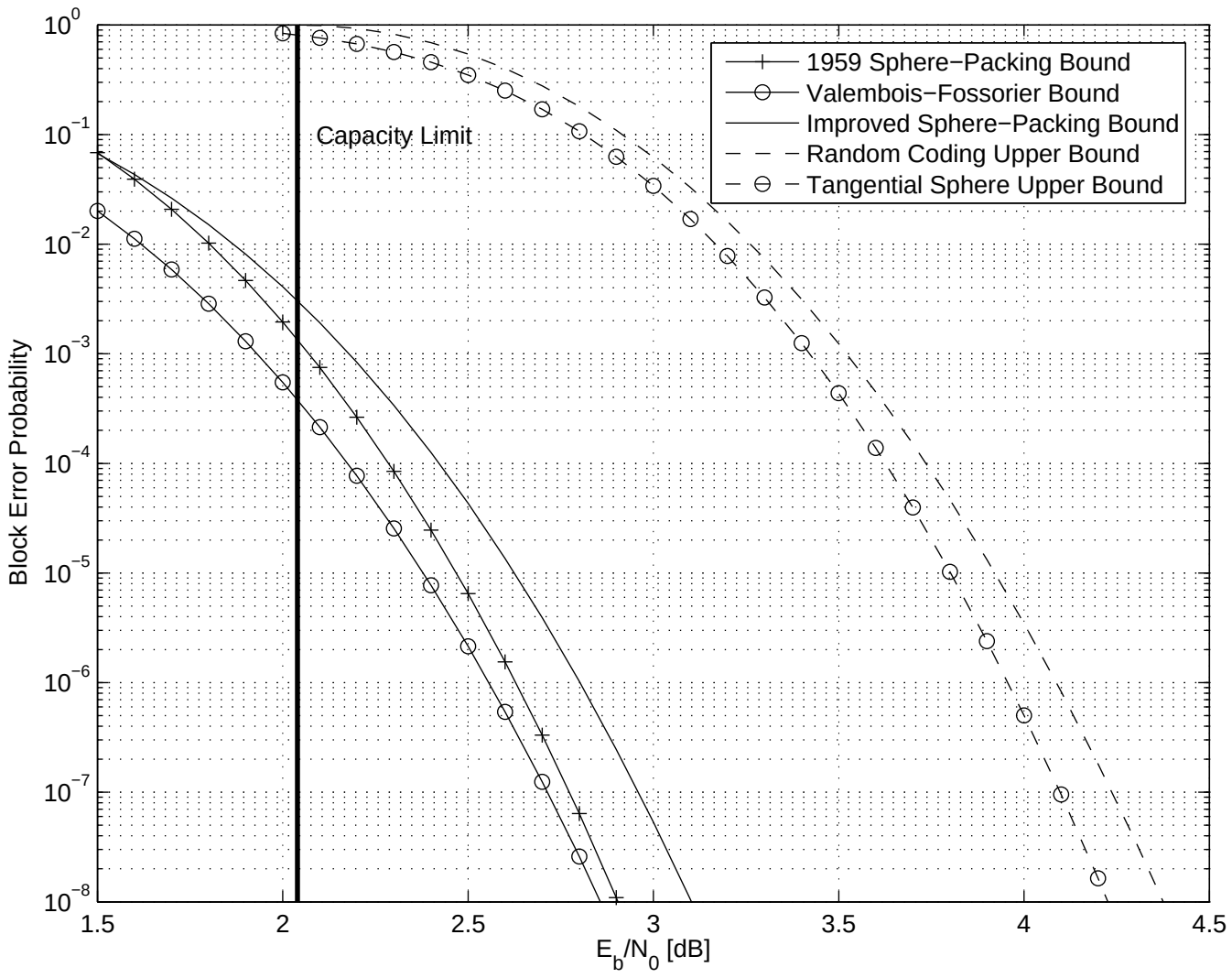

Fig. 1. A comparison between upper and lower bounds on the ML decoding error probability for block codes of length $N = 500$ bits and code rate of $0.8 \frac{\text{bits}}{\text{channel use}}$. This figure refers to BPSK modulated signals whose transmission takes place over an AWGN channel. The compared bounds are the 1959 sphere-packing (SP59) bound of Shannon [10], the Valembois-Fossorier (VF) bound [13], the improved sphere-packing (ISP) bound derived in Section III, the random-coding upper bound of Gallager [3], and the TSB [4], [7] when applied to fully random block codes with the above block length and rate.

Fig. 1 presents a comparison of the SP59 bound [10], the VF bound [13], and the ISP bound derived in Section III. The comparison refers to block codes of length 500 bits and rate $0.8 \frac{\text{bits}}{\text{channel use}}$ which are BPSK modulated and transmitted over an AWGN channel. The plot also depicts the random-coding bound of Gallager [3], the TSB [4], [7], and the capacity limit bound (CLB).[1] It is observed from this figure that even for relatively short block lengths, the ISP bound outperforms the SP59 bound for block error probabilities below $10^{-1}$. For a block error probability of $10^{-5}$, the ISP bound provides gains of about $0.2$ and $0.3$ dB over the SP59 bound and the VF bound, respectively. For these code parameters, the TSB provides a tighter upper bound on the block error probability of random codes than the random-coding bound; e.g., the gain of the TSB over the Gallager bound is about 0.2 dB for a block error probability of $10^{-5}$. Note that the Gallager bound is tighter than the TSB for fully random block codes of large enough block lengths, as the latter bound does not reproduce the random-coding error exponent for the AWGN channel [7]. However, Fig. 1 exemplifies the advantage of the TSB over the Gallager bound, when applied to random block codes of relatively short block lengths; this advantage is especially pronounced for low code rates where the gap between the error exponents of these two bounds is reduced (see [8, p. 67]), but it is also reflected from Fig. 1 for BPSK modulation with a code rate of $0.8 \frac{\text{bits}}{\text{channel use}}$. The gap between the TSB and the ISP bound, as upper and lower bounds respectively, is less than $1.2$ dB for all block error probabilities lower than $10^{-1}$. Also, the ISP bound is more informative than the CLB for block error probabilities below $3 \times 10^{-3}$.

[1]Although the CLB refers to the asymptotic case where the block length tends to infinity, it is plotted in [13] and here as a reference, in order to examine whether the improvement in the tightness of the ISP is for rates above or below capacity.

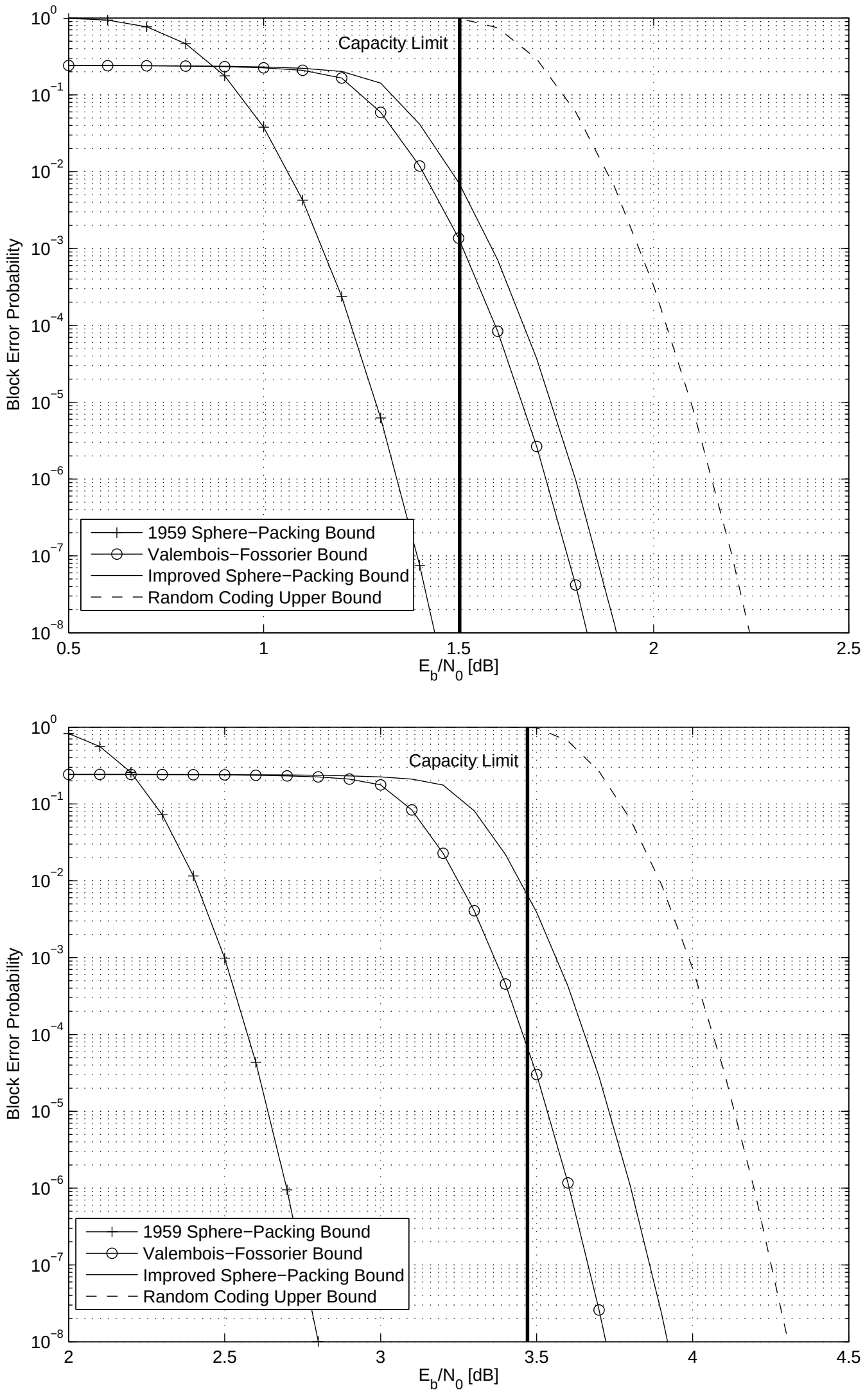

Fig. 2. A comparison of upper and lower bounds on the ML decoding error probability for block codes of length $N = 5580$ bits and information block length of 4092 bits. This figure refers to QPSK (upper plot) and 8-PSK (lower plot) modulated signals whose transmission takes place over an AWGN channel; the rates in this case are 1.467 and $2.200 \frac{\text{bits}}{\text{channel use}}$, respectively. The compared bounds are the 1959 sphere-packing (SP59) bound of Shannon [10], the Valembois-Fossorier (VF) bound [13], the improved sphere-packing (ISP) bound, and the random-coding upper bound of Gallager [3].

Fig. 2 presents a comparison of the bounds for codes of block length 5580 bits and 4092 information bits, where both QPSK (upper plot) and 8-PSK (lower plot) constellations are considered. The modulated signals correspond to 2790 and 1680 symbols, respectively, so the code rates for these constellations are 1.467 and 2.2 bits per channel use, respectively. For both constellations, the two considered SP67-based bounds (i.e., the VF and ISP bounds) outperform the SP59 for all block error probabilities below $10^{-1}$; the ISP bound provides gains of 0.1 and 0.2 dB over the VF bound for the QPSK and 8-PSK constellations, respectively. For both modulations, the gap between the ISP lower bound and the random-coding upper bound of Gallager does not exceed $0.4$ dB. In [1], Divsalar and Dolinar design codes with the considered parameters by using concatenated Hamming and accumulate codes. They also present computer simulations of the performance of these codes under iterative decoding, when the transmission takes place over the AWGN and several common modulation schemes are applied. For an error probability of $10^{-4}$, the gap between the simulated performance of these codes under iterative decoding, and the ISP lower bound, which gives an ultimate lower bound on the error probability of optimally designed codes under ML decoding, is approximately $1.4$ dB for QPSK and $1.6$ dB for 8-PSK signaling. This provides an indication on the performance of codes defined on graphs and their iterative decoding algorithms, especially in light of the feasible complexity of the decoding algorithm which is linear in the block length. To conclude, it is reflected from the results plotted in Fig. 2 that a gap of about 1.5 dB between the ISP lower bound and the performance of the iteratively decoded codes in [1] is mainly due to the imperfectness of these codes and their sub-optimal iterative decoding algorithm; this conclusion follows in light of the fact that for random codes of the same block length and rate, the gap between the ISP bound and the random coding bound is reduced to less than 0.4 dB.

While it was shown in Section III that the ISP bound is uniformly tighter than the VF bound (which in turn is uniformly tighter than the SP67 bound [11]), no such relations are shown between the SP59 bound and the recent improvements on the SP67 bound (i.e., the VF and ISP bounds). Fig. 3 presents regions of code rates and block lengths for which the ISP bound outperforms the SP59 bound and the CLB; it refers to BPSK modulated signals transmitted over the AWGN and considers block error probabilities of $10^{-4}$, $10^{-5}$ and $10^{-6}$. It is reflected from this figure that for any rate $0 < R < 1$, there exists a block length $N = N(R)$ such that the ISP bound outperforms the SP59 bound for block lengths larger than $N(R)$; the same property also holds for the VF bound, but the value of $N(R)$ depends on the considered SP67-based bound, and it becomes significantly larger in the comparison of the VF and SP59 bounds. It is also observed that the value $N(R)$ is monotonically decreasing with $R$, and it approaches infinity as we let $R$ tend to zero. An intuitive explanation for this behavior can be given by considering the capacity limits of the binary-input and the energy-constraint AWGN channels. For any value $0 \leq C < 1$, denote by $\frac{E_{\mathrm{b},1}(C)}{N_0}$ and $\frac{E_{\mathrm{b},2}(C)}{N_0}$ the values of $\frac{E_{\mathrm{b}}}{N_0}$ required to achieve a channel capacity of $C$ bits per channel use for the binary-input and energy-constraint AWGN channels, respectively (note that in the latter case, the input distribution which achieves capacity is also Gaussian). For any $0 \leq C < 1$, clearly $\frac{E_{\mathrm{b},1}(C)}{N_0} \geq \frac{E_{\mathrm{b},2}(C)}{N_0}$; however, the difference between these values is monotonically increasing with the capacity $C$, and, on the other hand, this difference approaches zero as we let $C$ tend to zero. Since the SP59 bound only constrains the signals to be of equal energy, it gives a measure of performance for the energy-constraint AWGN channel, where the SP67-based bounds consider the actual modulation and therefore refer to the binary-input AWGN channel. As the code rates become higher, the difference in the ultimate performance between the two channels is larger, and therefore the SP67-based bounding techniques outperform the SP59 bound for smaller block lengths. For low code rates, the difference between the channels is smaller, and the SP59 outperforms the SP67-based bounding techniques even for larger block lengths due to the superior bounding technique which is specifically tailored for the AWGN channel. Fig 4 presents the regions of code rates and block lengths for which the VF bound (upper plot) and the ISP bound (lower plot) outperform the CLB and the SP59 bound when the signals are BPSK modulated and transmitted over the AWGN channel; block error probabilities of $10^{-4}$, $10^{-5}$ and $10^{-6}$ are examined. This figure is focused on high code rates, where the performance of the SP67-based bounds and their advantage over the SP59 bound is most appealing. From Figure 4, we have that for a code rate of 0.75 bits per channel use and an error probability of $10^{-6}$, the VF bound becomes tighter than the SP59 for block lengths exceeding 870 bits while the ISP bound reduces this value to 617 bits; moreover, when increasing the rate to 0.8 bits per channel use, the respective minimal block lengths reduce to 550 and 350 bits for the VF and ISP bounds, respectively. Fig 5 shows the regions of code rates and block lengths where the ISP outperforms the CLB and SP59 bounds for QPSK (upper plot) and 8-PSK (lower plot) modulations. Comparing the lower plot of Fig. 4 which refers to BPSK modulation with the upper plot of Fig. 5 which refers to QPSK modulation, one can see that the two graphs are virtually identical (when accounting for the doubling of the rate which is due to the use of both real and imaginary dimensions in the QPSK modulation). This is due to the fact that QPSK modulation poses no additional constraints on the channel and in fact, the real and imaginary planes can be serialized and decoded as in BPSK modulation. However, this property does not hold when replacing the ISP bound by the VF bound; this is due to the fact that the VF bound considers a fixed composition subcode of the original code and the increased size of the alphabet causes a greater loss in the rate for QPSK modulation. When comparing the two plots of Fig. 5, it is evident that the minimal value of the block length for which the ISP bound becomes better than the SP59 bound decreases as the size of the input alphabet is increased (when the rate is given in units of information bits per code bit). An intuitive justification for this phenomenon is attributed to the fact that referring to the constellation points of the M-ary PSK modulation, the mutual information between the code symbols in each dimension of the QPSK modulation is zero, while as the spectral efficiency of the PSK modulation is increased, the mutual information between the real and imaginary parts of each signal point is increased; thus, as the spectral efficiency is increased, this poses a stronger constraint on the possible positioning of the equal-energy signal points on the $N$-dimensional sphere. This intuition suggests an explanation for the reason why as the spectral efficiency is increased, the advantage of the ISP bound over the SP59 bound (where the latter does not take into account the modulation scheme) holds even for smaller block lengths. This effect is expected to be more subtle for the VF bound since a larger size of the input alphabet decreases the rate for which the error exponent is evaluated (see (11)).

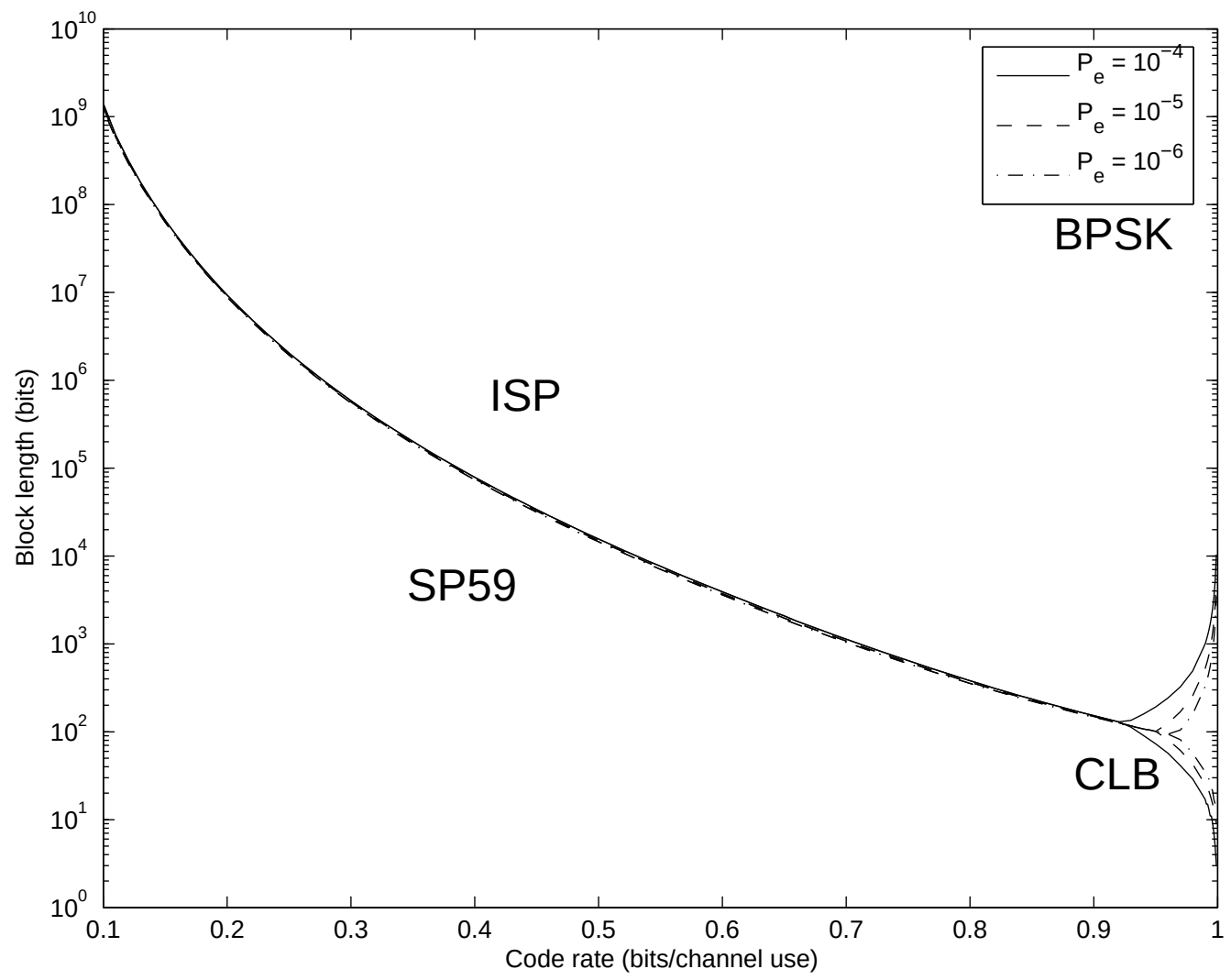

Fig. 3. Regions in the two-dimensional space of code rate and block length, where a bound is better than the two others for three different targets of block error probability ($P_{\mathrm{e}}$). The figure compares the tightness of the 1959 sphere-packing (SP59) bound of Shannon [10], the improved sphere-packing (ISP) bound, and the capacity-limit bound (CLB). The plot refers to BPSK modulated signals whose transmission takes place over the AWGN channel, and the considered code rates lie in the range between 0.1 and $1 \frac{\text{bits}}{\text{channel use}}$.

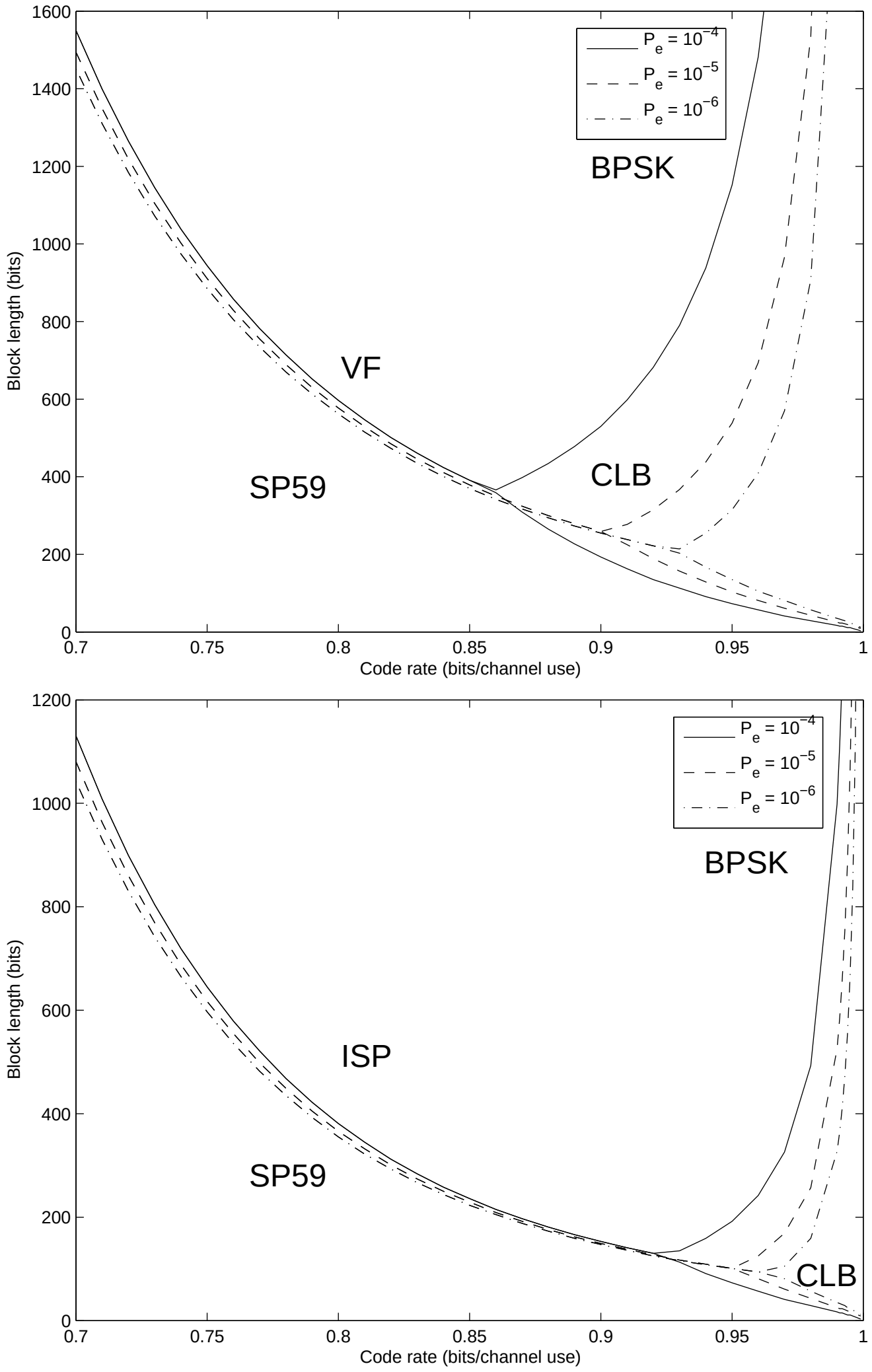

Fig. 4. Regions in the two-dimensional space of code rate and block length, where a bound is better than the two others for three different targets of block error probability ($P_{\mathrm{e}}$). The figure compares the tightness of the 1959 sphere-packing (SP59) bound of Shannon [10], the capacity-limit bound (CLB), and the Valembois-Fossorier (VF) bound [13] (upper plot) or the improved sphere-packing (ISP) bound in Section III (lower plot). The plots refer to BPSK modulated signals whose transmission takes place over the AWGN channel, and the considered code rates lie in the range between 0.70 and $1\frac{\text{bits}}{\text{channel use}}$.

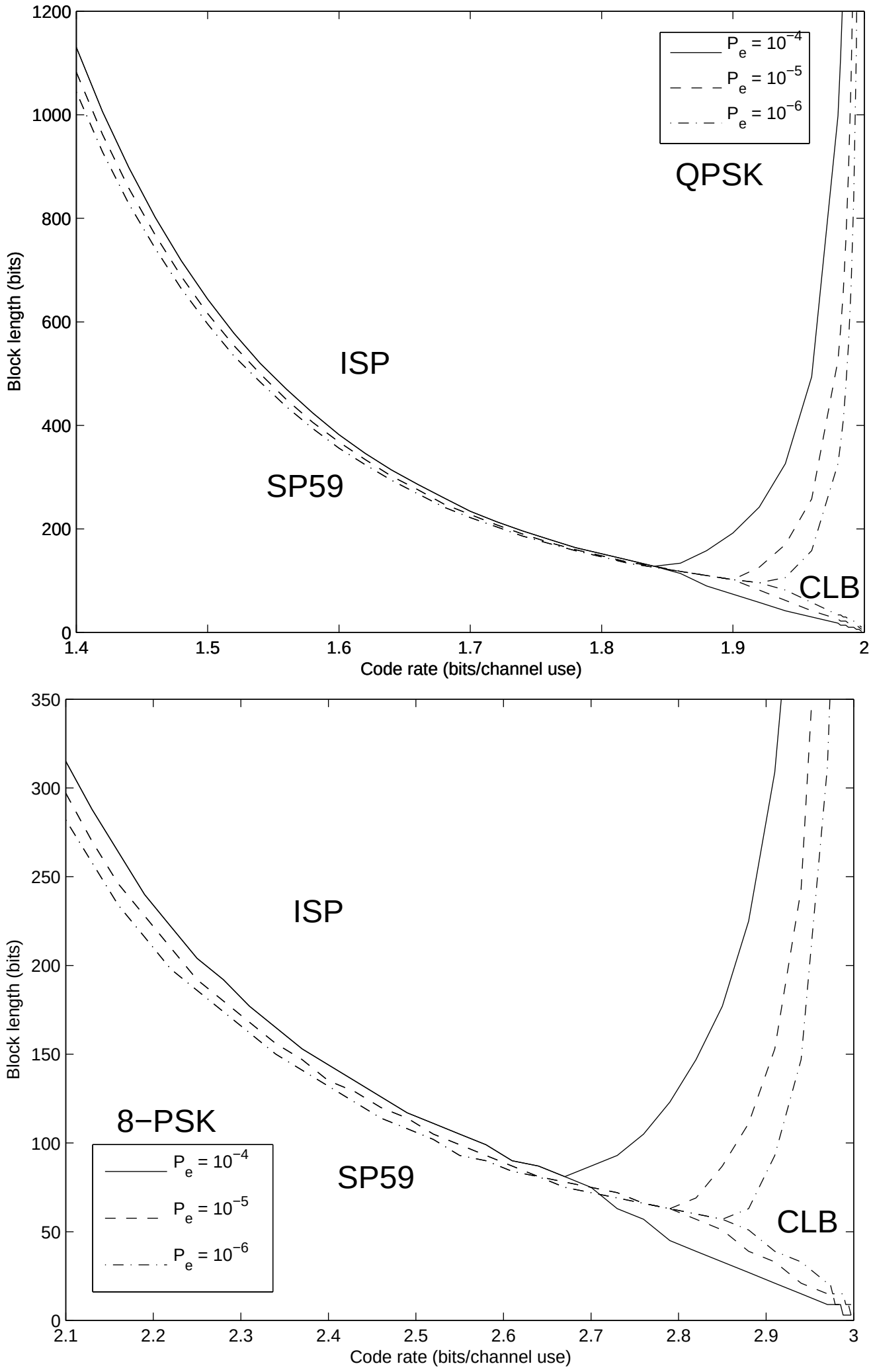

Fig. 5. Regions in the two-dimensional space of code rate and block length, where a bound is better than the two others for three different targets of block error probability ($P_{\mathrm{e}}$). The figure compares the tightness of the 1959 sphere-packing (SP59) bound of Shannon [10], the improved sphere-packing (ISP) bound, and the capacity-limit bound (CLB). The plots refer to QPSK (upper plot) and 8-PSK (lower plot) modulated signals whose transmission takes place over the AWGN channel; the considered code rates lie in the range between 1.4 and $2\frac{\text{bits}}{\text{channel use}}$ for the QPSK modulated signals and between 2.1 and $3\frac{\text{bits}}{\text{channel use}}$ for the 8-PSK modulated signals.

## V. SUMMARY

This paper presents an improved sphere-packing (ISP) bound targeting codes of short to moderate block lengths, and exemplifies some of its applications. The derivation of the ISP bound was stimulated by the remarkable performance and feasible complexity of turbo-like codes with short to moderate block lengths. We first review the 1967 sphere-packing (SP67) bound for discrete memoryless channels [11], and a recent improved bound derived by Valembois and Fossorier [13]. The ISP bound relies on concepts used for their derivation, and provides a uniformly tighter lower bound on the decoding error probability. Under a mild condition, the validity of the ISP bound is extended to general memoryless channels (even with continuous input and output alphabets); the basic observation which enables the derivation of the ISP bound is explained in Remark III.1 (see p. 8).

We apply the ISP bound to M-ary PSK block coded modulation schemes whose transmission takes place over the AWGN channel and the received signals are coherently detected. The tightness of the ISP bound is exemplified by comparing it with upper and lower bounds on the ML decoding error probability. It is shown that the ISP bound suggests an interesting alternative to the 1959 sphere-packing (SP59) bound of Shannon [10], where the latter is specialized for the AWGN channel. The full paper version [15] also presents numerical results for the binary erasure channel (BEC); the exact performance of fully random binary linear block codes under ML decoding is compared with the ISP bound when the transmission takes place over the BEC.

As evidenced in the literature (see, e.g., [2], [5], [8], [13]), the sphere-packing bounds are useful for assessing the theoretical limitations of block codes and the power of iteratively decoded codes.

The ISP bound is especially attractive for block codes of high rate in terms of the range of the block lengths where this bound outperforms the SP59 and capacity limit bounds (see Figs. 2–5). Its improvement over the SP67 bound and the bound in [13, Theorem 7] is especially pronounced as the input alphabet of the considered modulation is increased.

This paper presents in part the results introduced in the full paper version [15]. Part of the work in [15] which is not presented here is related to a discussion on the SP59 bound and a derivation of an algorithm which enables the numerical computation of the SP59 bound in the log-domain; this is in contrast to expressions introduced in [10] and [13] where the SP59 bound is calculated in the probability domain and its calculation is subject to underflows and overflows even for relatively short block lengths. The new algorithm enables the exact calculation of the SP59 bound for moderate to large block lengths without resorting to the asymptotic approximations in [10]. This algorithmic issue is addressed in [15, Section 4] and the conference paper [9].